\documentclass[aps,pra,showpacs,superscriptaddress,preprint]{revtex4}
\usepackage{amsmath}
\usepackage{graphicx}

\catcode`ð=\active
 \defð{\u{g}}
 \catcode`Ð=\active
 \defÐ{\u{G}}
 \catcode`Ý=\active
\defÝ{\. I}
 \catcode`ö=\active
\defö{\"{o}}
 \catcode`Ö=\active
 \defÖ{\"O}
 \catcode`ü=\active
 \defü{\"{u}}
 \catcode`Ü=\active
 \defÜ{\"{U}}
 \catcode`Þ=\active
\defÞ{\c{S}}
 \catcode`þ=\active
 \defþ{\c{s}}
 \catcode`ý=\active
 \defý{{\i}}
 \catcode`ç=\active
\defç{\d{c}}
 \catcode`Ç=\active
\defÇ{\d{C}}

\input{tcilatex}

\begin{document}

\title{\textbf{Quantization rule solution to the Hulth\'{e}n potential in
arbitrary dimension by a new approximate scheme for the centrifugal term }}
\author{Sameer M. Ikhdair}
\email[E-mail: ]{sikhdair@neu.edu.tr}
\affiliation{Physics Department, Near East University, Nicosia, North Cyprus, Turkey}
\date{%
\today%
}

\begin{abstract}
The bound state energies and wave functions for a particle exposed to the
Hulth\'{e}n potential field in the $D$-dimensional space are obtained within
the improved quantization rule for any arbitrary $l$ state. The present
approximation scheme used to deal with the centrifugal term in the effective
Hulth\'{e}n potential is systematic and accurate. The solutions for the
three-dimensional ($D=3$) case and the $s$-wave ($l=0$) case are briefly
discussed.

\end{abstract}

\maketitle

\newpage

\section{Introduction}

It is well-known that the exact analytical solution of the hyperradial Schr%
\"{o}dinger equation, in any arbitrary spatial dimension $(D\geq 2),$ for
its bound state energy levels, is fundamental in understanding the bound
energy spectrum of nonrelativistic and relativistic quantum mechanics, since
the resulting wave function contains all the necessary information to
describe fully any quantum system. There are only a few potentials for which
the radial Schr\"{o}dinger equation can be solved explicitly for all $n$ and 
$l$ quantum numbers$.$ One of these exactly solvable potentials is the Hulth%
\'{e}n potential [1,2] which can be solved in a closed form for $s$ wave ($%
l=0$). However, the three-dimensional radial Schr\"{o}dinger equation for
the spherically symmetric Hulth\'{e}n potential cannot be solved
analytically for $l\neq 0$ states because of the centrifugal term $\sim
r^{-2}$ [3-5]$.$ The Hulth\'{e}n potential is one of the important molecular
potentials used in different areas of physics to describe the interaction
between two atoms and has attracted a great of interest for some decades in
the history of quantum chemistry. Until now, it has also been used
extensively to describe the molecular structure and a possible form of
atomic and nuclear interactions [6-10]. \ Further, the study of this
potential is only a special example of those exponential type potentials
[11] such as the modified hyperbolic-type potentials (Scarf, modified
Rosen-Morse, modified second type P\"{o}schl-Teller) [12], Eckart [13],
Rosen-Morse [14], Manning-Rosen [15] potentials and so forth. So far,
numerous attempts have been developed to calculate the bound-state energies
such as variational [10], supersymmetry quantum mechanics [16,17],
Nikiforov-Uvarov method (NU) [5,14,18,19], asymptotic iteration method (AIM)
[20], hypervirial perturbation [21], shifted $1/N$ expansion (SE) and
modified shifted $1/N$ expansion (MSE) [22], exact (improved) quantization
rule (EQR or IQR) [23,24], perturbative formalism [25-29], polynomial
solution [30], wave function ansatz method [31], factorization method [32]
and tridiagonal J--matrix representation (TJM) [33] which split original
Hamiltonian into two parts as $H=H_{0}+V$ where $H_{0}$ is the part of the
Hamiltonian that could be treated analytically while the remaining part, $V,$
has to be treated numerically.to solve the radial Schr\"{o}dinger,
Klein-Gordon and Dirac equations exactly or quasi-exactly for $l\neq 0$
within a given potential.

Recently, Ma and Xu have proposed an exact (improved) quantization rule (EQR
or IQR) and shown its power in calculating the energy levels of all bound
states for some solvable quantum systems [23,24]. The method has been shown
to be effective for calculating the bound state solutions of the Schr\"{o}%
dinger and Dirac wave equations with a spherically symmetric potential
[34-39]. So far, it has been applied, with great success, to study a great
number of potentials like the rotating Morse [34,35], the Kratzer-type [36],
the trigonometric Rosen-Morse [37], the hyperbolic and the second P\"{o}%
schl-Teller-like potentialas [38], the Hulth\'{e}n potential [39] and the
Woods-Saxon potential [40] and so forth. Very recently, Gu and Sun [39] have
extended the application of the IQR to the solution of the $D$-dimensional
Schr\"{o}dinger equation with the Hulth\'{e}n potential for $l\neq 0$ using
the usual approximation to deal with the centrifugal term [41-43].

Very recently, Dong [12] has introduced a more beautiful exact quantization
rule to simplify the calculation of the energy levels for exactly solvable
quantum systems. The energy spectra of the modified hyperbolic-type
potentials have been carried out by this rule. Qiang and Dong [44] have
found a proper quantization rule (PQR) and showed that the previous
complicated and tedious calculations for the energy spectra can be greatly
simplified. This new quantization rule can be applied to any exactly
solvable potential. Qiang-Dong PQR has been further applied to exactly
solvable shape invariant potentials [45]. Very recently, Yin \textit{et al.}
[46] have shown that the SWKB is exact for all shape invariant potentials
(SIPs).

In this paper, we aim to extend the study of Ref. [39] by using a new
improved approximate scheme to deal with the centrifugal term. Further, we
solve the present potential model on the assumption that the space may
posses an arbitrary number of spatial dimensions $D.$ This arbitrary
dimensional study enables one to give analytical tests using energy
calculations for interdimensional degeneracy, i.e., $(n,l,D)\rightarrow
(n,l\pm 1,D\mp 2)$ corresponding to the confined $D=2-4$ dimensional Hulth%
\'{e}n potential.

It is worth noting that this alternative approximating approach has shown
its accuracy in calculating the analytic and numerical energy spectrum of
the Hulth\'{e}n potential for $l\neq 0$ [3-5]. Further, it has also been
applied to the spin and pseudospin symmetries, e.g., Wei and Dong have
studied the approximation of the Dirac equation with scalar and vector
modified and deformed generalized P\"{o}schl-Teller and Manning-Rosen
potentials within the improved approximation formula to the centrifugal term
[47-50].

This paper is organized as follows. In Sec. 2, the EQR (IQR) method is
reviewed and extended to any arbitrary dimension $(D\geq 2)$. In Sec. 3, the 
$D$-dimensional ($D\geq 2$) Schr\"{o}dinger equation is solved by this
method with $l\neq 0$ quantum numbers to obtain the energy eigenvalues. In
Sec. 4, we calculate the corresponding hyperradial wave functions of the
Hulth\'{e}n potential. Finally, some conclusions are given in Sec. 5.

\section{Exact (Improved) Quantization Rule}

A brief outline to the improved quantization rule is presented with an
extension to the $D$-dimensional space ($D\geq 2$). The details can be found
in Refs. [23,24]. The IQR has recently been proposed to solve exactly the
one-dimensional ($1D$) Schr\"{o}dinger equation:

\begin{equation}
\psi {}^{\prime \prime }(x)+k(x)^{2}\psi (x)=0,\text{ \ }k(x)=\frac{\sqrt{%
2\mu \left[ E-V(x)\right] }}{\hbar },
\end{equation}%
where the prime denotes the derivative with respect to the variable $x.$
Here $\mu $ is the reduced mass of the two interacting particles, $k(x)$ is
the momentum and $V(x)$ is a piecewise continuous real potential function of 
$x.$ The Schr\"{o}dinger equation is equivalent to the Riccati equation%
\begin{equation}
\phi {}^{\prime }(x)+\phi (x)^{2}+k(x)^{2}=0,
\end{equation}%
where $\phi (x)=\psi {}^{\prime }(x)/\psi (x)$ is the logarithmic derivative
of wave function $\psi (x).$ Due to the Sturm-Liouville theorem, the $\phi
(x)$ decreases monotonically with respect to $x$ between two turning points,
where $E\geq V(x).$ Specifically, as $x$ increases across a node of the wave
function $\psi (x),$ $\phi (x)$ decreases to $-\infty ,$ jumps to $+\infty ,$
and then decreases again.

Moreover, Ma and Xu [23,24] have generalized this exact quantization rule to
the three-dimensional $\left( 3D\right) $ radial Schr\"{o}dinger equation
with spherically symmetric potential by simply making the replacements $%
x\rightarrow $ $r$ and $V(x)\rightarrow V_{\text{eff}}(r)$:

\begin{equation}
\int\limits_{r_{A}}^{r_{B}}k(r)dr=N\pi
+\int\limits_{r_{A}}^{r_{B}}k{}^{\prime }(r)\frac{\phi (r)}{\phi {}^{\prime
}(r)}dr,\text{ }k(r)=\frac{\sqrt{2\mu \left[ E_{n,l}-V_{\text{eff}}(r)\right]
}}{\hbar },
\end{equation}%
where $r_{A}$ and $r_{B}$ are two turning points determined from the
relation $E_{n,l}=V_{\text{eff}}(r),$ $N=n+1$ is the number of nodes of \ $%
\phi (r)$ in the region $E_{n,l}\geq V_{\text{eff}}(r)$ and it is larger by
one than the number of nodes of wave function $\psi (r).$ The first term $%
N\pi $ is the contribution from the nodes of the logarithmic derivative of
wave function, and the second term in (3) is called the quantum correction.
It is found that, for all well-known exactly solvable quantum systems, this
quantum correction is independent of the number of nodes of wave function of
the system. This means that it is enough to consider the ground state in
calculating the quantum correction, i.e., 
\begin{equation}
Q_{c}=\int\limits_{r_{A}}^{r_{B}}k_{0}{}^{\prime }(r)\frac{\phi _{0}(r)}{%
\phi _{0}{}^{\prime }(r)}dr=\pi q,
\end{equation}%
The quantization rule still holds for Schr\"{o}dinger equation with
spherically symmetric potential in $D$ dimensions. In what follows, we shall
employ this method to extend the work of Ref. [39] by using an improved
approximation to the centrifugal term.

\section{Eigenvalues of the Hulth\'{e}n potential}

The Schr\"{o}dinger equation with spherically symmetric potential $V(r)$ for 
$l\neq 0$ takes the simple form 
\begin{equation}
\left( -\frac{\hbar ^{2}}{2\mu }\nabla _{D}^{2}+V(r)-E_{n,l}\right) \psi
_{n,l,m}(r,\Omega _{D})=0,\text{ }
\end{equation}%
where the representation of the Laplacian operator $\nabla _{D}^{2},$ in
spherical coordinates, is 
\begin{equation}
\nabla _{D}^{2}=\frac{\partial ^{2}}{\partial r^{2}}+\frac{\left( D-1\right) 
}{r}\frac{\partial }{\partial r}-\frac{l\left( l+D-2\right) }{r^{2}},
\end{equation}%
and%
\begin{equation}
\psi _{n,l,m}(r,\Omega _{D})=\psi _{n,l}(r)Y_{l}^{m}(\Omega _{D}),\text{ }%
\psi _{n,l}(r)=r^{-(D-1)/2}R(r),
\end{equation}%
where $Y_{l}^{m}(\Omega _{D})$ is the hyperspherical harmonics. The wave
functions $\psi _{n,l,m}(r,\Omega _{D})$ belong to the energy eigenvalues $%
E_{n,l}$ and $V(r)$ stands for the Hulth\'{e}n potential in the
configuration space and $r$ represents the $D$-dimensional intermolecular
distance $\left( \dsum\limits_{i=1}^{D}x_{i}^{2}\right) ^{1/2}.$

Further, substituting Eqs. (6) and (7) into Eq. (5) yields the wave equation
satisfying the radial wave function $R(r)$ in a simple analogy to the $2D$
and $3D$ radial Schr\"{o}dinger equation 
\begin{equation}
R^{\prime \prime }(r)+\frac{2\mu }{\hbar ^{2}}\left[ E_{n,l}-V_{eff}(r)%
\right] R(r)=0,
\end{equation}%
where $V_{eff}(r)$ is the Hulth\'{e}n effective potential in $D$ dimensions
defined by 
\begin{equation}
V_{eff}(r)=-Ze^{2}\alpha \frac{e^{-\alpha r}}{1-e^{-\alpha r}}+\frac{\left(
\Lambda ^{2}-1\right) \hbar ^{2}}{8\mu r^{2}},
\end{equation}%
with the parameter 
\begin{equation}
\Lambda =2l+D-2.
\end{equation}%
The radial wave function $R(r)$ satisfying Eq. (8) should be normalizable
and finite near $r=0$ and $r\rightarrow \infty $ for the bound-state
solutions. The wave equation (8) with the Hulth\'{e}n potential is an
exactly solvable problem for $l=0$ ($s$-wave) [2,51-53], however, it cannot
be solved analytically when $l\neq 0$ because of the centrifugal barrier
term, i.e., $\left( \Lambda ^{2}-1\right) \hbar ^{2}r^{-2}/(8\mu )$.
Therefore, to solve Eq. (8) analytically, we must use a new approximation
scheme of the exponential-type proposed recently by Jia \textit{et al} (cf.
e.g., [54-59]) to deal with the centrifugal term:%
\begin{equation}
\frac{1}{r^{2}}\approx \alpha ^{2}\left( c_{0}+\frac{e^{-\alpha r}}{\left(
1-e^{-\alpha r}\right) ^{2}}\right) ,
\end{equation}%
where the dimensionless constant $c_{0}=1/12$ is exact as reported by other
authors (cf. e.g., [3-5])$.$ Very recently, we have applied the above
approximation scheme (11) to obtain improved bound state solutions to the
Schr\"{o}dinger equation with the Manning-Rosen potential for arbitrary $l$%
-waves [15]. Obviously, the above approximation to the centrifugal term
turns to $r^{-2}$ when the parameter $\alpha $ goes to zero (small screening
parameter $\alpha $) as%
\begin{equation}
\underset{\alpha \rightarrow 0}{\lim }\left[ \alpha ^{2}\left( c_{0}+\frac{1%
}{e^{\alpha r}-1}+\frac{1}{\left( e^{\alpha r}-1\right) ^{2}}\right) \right]
=\frac{1}{r^{2}},  \tag{11a}
\end{equation}%
which shows that the usual approximation is the limit of our approximation
(cf. e.g., [4] and the references therein). Further, by defining%
\begin{equation}
a=Ze^{2}\alpha ,\text{ }b=\alpha ^{2}L^{2},\text{ }L^{2}=\frac{\hbar ^{2}}{%
2\mu }\left( l+\frac{D-1}{2}\right) \left( l+\frac{D-3}{2}\right) ,
\end{equation}%
then we have from Eq. (8):%
\begin{equation}
R^{\prime \prime }(r)+\frac{2\mu }{\hbar ^{2}}\left[ E_{n,l}+a\frac{%
e^{-\alpha r}}{1-e^{-\alpha r}}-b\left( c_{0}+\frac{e^{-\alpha r}}{\left(
1-e^{-\alpha r}\right) ^{2}}\right) \right] R(r)=0,\text{ }
\end{equation}%
where $E_{n,l}$ is the bound state energy of the system and $n$ and $l$
signify the radial and angular quantum numbers, respectively.

We now study this system through the improved exact quantization rule. At
first, we introduce a new variable%
\begin{equation}
z(r)=\frac{e^{-\alpha r}}{1-e^{-\alpha r}},\text{ }z{}^{\prime }(r)=-\alpha
z(1+z),
\end{equation}%
where $r\in (0,\infty )$ and $z\in (0,\infty ).$ Overmore, the turning
points $z_{A\text{ }}$ and $z_{B\text{ }}$ are determined by solving $%
V_{eff}(z)=bz^{2}+\left( b-a\right) z+bc_{0}=E_{n,l}$ as follows:%
\begin{equation}
z_{A\text{ }}=\frac{a}{2b}-\frac{1}{2}-\frac{1}{2b}\sqrt{\left( a-b\right)
^{2}+4b\left( E_{nl}-bc_{0}\right) },\text{ }z_{B\text{ }}=\frac{a}{2b}-%
\frac{1}{2}+\frac{1}{2b}\sqrt{\left( a-b\right) ^{2}+4b\left(
E_{n,l}-bc_{0}\right) },
\end{equation}%
with the properties%
\begin{equation}
z_{A\text{ }}+z_{B\text{ }}=\frac{a}{b}-1,\text{ }z_{A\text{ }}z_{B\text{ }%
}=-\frac{E_{nl}}{b}+c_{0}.
\end{equation}%
The momentum $k(z)$ between two turning points is expressed as 
\begin{subequations}
\begin{equation}
k(z)=\frac{\sqrt{2\mu b}}{\hbar }\sqrt{-z^{2}+\frac{\left( a-b\right) }{b}z-%
\frac{\left( bc_{0}-E_{n,l}\right) }{b}}=\frac{\sqrt{2\mu b}}{\hbar }\sqrt{%
\left( z_{B\text{ }}-z\right) \left( z-z_{A\text{ }}\right) },
\end{equation}%
\begin{equation}
\frac{dk(z)}{dz}=\frac{\sqrt{2\mu b}}{2\hbar }\left( \sqrt{\frac{z_{B\text{ }%
}-z}{z-z_{A\text{ }}}}-\sqrt{\frac{z-z_{A\text{ }}}{z_{B\text{ }}-z}}\right)
,
\end{equation}%
The Riccati equation (2) now becomes 
\end{subequations}
\begin{equation}
-\alpha z\left( z+1\right) \frac{d\phi _{0}(z)}{dz}=-\frac{2\mu }{\hbar ^{2}}%
\left[ E_{0}-bz^{2}+\left( a-b\right) z-bc_{0}\right] -\phi _{0}(z)^{2},
\end{equation}%
having the only possible solution satisfying%
\begin{equation}
\phi _{0}(r)=c_{1}z+c_{2},\text{ }\phi _{0}{}^{\prime }(r)=-\alpha
c_{1}z(1+z),\text{ }c_{1}>0.
\end{equation}%
where we have used $\phi _{0}(r)\equiv \phi _{0}(z).$ Substituting $\phi
_{0}(z)$ into Eq. (18), one has the ground state energy eigenvalue and wave
function solutions%
\begin{equation}
\left\{ 
\begin{array}{c}
\phi _{0}(z)=\nu \alpha z+\alpha \left( \frac{\mu Ze^{2}}{\hbar ^{2}}\frac{1%
}{\nu \alpha }-\frac{\nu }{2}\right) , \\ 
\nu =\frac{1}{2}+\frac{1}{2}\sqrt{1+\frac{8\mu }{\hbar ^{2}}L^{2}}=l+\frac{%
D-1}{2},\text{ }\nu \geq 1, \\ 
\widetilde{E}_{n=0}=E_{0}-bc_{0}=-\frac{\hbar ^{2}\alpha ^{2}}{2\mu }\left( 
\frac{\mu Ze^{2}}{\hbar ^{2}}\frac{1}{\nu \alpha }-\frac{\nu }{2}\right)
^{2}.%
\end{array}%
\right.
\end{equation}%
After a lengthy algebra but straightforward, we can calculate the integral
of the quantum correction (4) based on the ground state as%
\begin{equation}
Q_{c}=\pi q=\pi \left( \frac{\sqrt{2\mu }}{\hbar }L+\nu -1\right) .
\end{equation}%
The integral of the momentum $k(r)$ in the quantization rule (3) is
calculated as%
\begin{equation*}
\int\limits_{r_{A}}^{r_{B}}k(r)dr=-\frac{\sqrt{2\mu b}}{\alpha \hbar }%
\int\limits_{z_{A}}^{z_{B}}\left( \frac{\sqrt{\left( z-z_{A\text{ }}\right)
\left( z_{B\text{ }}-z\right) }}{z}-\frac{\sqrt{\left( z-z_{A\text{ }%
}\right) \left( z_{B\text{ }}-z\right) }}{1+z}\right) dz
\end{equation*}%
\begin{equation}
=\frac{2\mu }{\hbar ^{2}}L\left( 1+\sqrt{c_{0}-\frac{E_{n,l}}{b}}-\sqrt{%
c_{0}+\frac{\left( a-E_{n,l}\right) }{b}}\right) \pi .
\end{equation}%
Using the relations (21) and (22), the improved quantization rule (3) turn
out to be%
\begin{equation}
\pi \frac{\sqrt{2\mu }}{\hbar }L\left( 1+\sqrt{c_{0}-\frac{E_{n,l}}{b}}-%
\sqrt{c_{0}+\frac{\left( a-E_{n,l}\right) }{b}}\right) =\pi \left( \frac{%
\sqrt{2\mu }}{\hbar }\Lambda +n+\nu \right) .
\end{equation}%
Thus, one can finally find the approximation to the bound state energy
levels $E_{nl}$ for the $D$-dimensional Hulth\'{e}n potential,%
\begin{equation}
E_{n,l}^{(D)}=\frac{\hbar ^{2}\alpha ^{2}}{2\mu }\left\{ \left( l+\frac{D-1}{%
2}\right) \left( l+\frac{D-3}{2}\right) c_{0}-\left[ \frac{\mu Ze^{2}}{\hbar
^{2}\left( n+l+\frac{D-1}{2}\right) \alpha }-\frac{\left( n+l+\frac{D-1}{2}%
\right) }{2}\right] ^{2}\right\} ,
\end{equation}%
where $n,l=0,1,2,\cdots .$ Therefore, the energy spectrum in $3D$ space can
be obtained as%
\begin{equation}
E_{n,l}=\frac{\hbar ^{2}\alpha ^{2}}{2\mu }\left\{ \frac{l\left( l+1\right) 
}{12}-\left[ \frac{\mu Ze^{2}}{\hbar ^{2}\left( n+l+1\right) \alpha }-\frac{%
\left( n+l+1\right) }{2}\right] ^{2}\right\} ,\text{ }n,l=0,1,2,\cdots ,
\end{equation}%
which is identical to Eq. (34) of Ref. [4]. In the case of the $s$-wave ($%
l=0 $), the previous relation turns out to become%
\begin{equation}
E_{n}=-\frac{\hbar ^{2}\alpha ^{2}}{2\mu }\left[ \frac{\mu Ze^{2}}{\hbar
^{2}\left( n+1\right) \alpha }-\frac{\left( n+1\right) }{2}\right] ^{2},%
\text{ }n=0,1,2,\cdots ,
\end{equation}%
which is identical to the ones obtained before using the AIM [20], SUSYQM
approach [60-64], quasi-linearization method [65] and NU method [42,66].
Overmore, if we take the dimensionless constant $c_{0}=0$ in the present
approximation, Eq. (24) reduces to%
\begin{equation}
E_{n,l}=-\frac{\hbar ^{2}\alpha ^{2}}{2\mu }\left[ \frac{\mu Ze^{2}}{\hbar
^{2}\left( n+l+\frac{D-1}{2}\right) \alpha }-\frac{\left( n+l+\frac{D-1}{2}%
\right) }{2}\right] ^{2},
\end{equation}%
which is consistent with the energy eigenvalues formula given in Eq. (32) of
Ref. [20], Eq. (24) of Ref. [67] and Eq. (28) of Ref. [42] when $D=3.$ By
taking the chosen parameters $\hbar =2\mu =e=1$ and for $Z=1,$ the above
result is consistent with Eq. (24) of Ref. [39]. The critical screening
parameter can be found as $\alpha _{c}=\frac{2\mu Ze^{2}}{\hbar ^{2}\left(
n+l+1\right) ^{2}}$ when $E_{nl}=0$ and $c_{0}=0.$

\section{Eigenfunctions}

We are now in the position to study the corresponding eigenfunction of this
quantum system for completeness. The Riccati equation of the relation (8) is
[68]%
\begin{equation}
\phi {}^{\prime }(r)=-\frac{2\mu }{\hbar ^{2}}\left[ E_{nl}-V_{eff}(r)\right]
-\phi (r)^{2},
\end{equation}%
where%
\begin{equation}
\phi (r)=\frac{R{}^{\prime }(r)}{R(r)}.
\end{equation}%
Based on 
\begin{equation}
R(r)=e^{\int^{r}\phi (r)dr}=e^{-\frac{1}{\alpha }\int^{r}\frac{1}{z(z+1)}%
\phi (z)dz},
\end{equation}%
and using Eq. (19), we can easily calculate the eigenfunction of the ground
state as%
\begin{equation}
R_{0}(r)=N_{0}\left( e^{-\alpha r}\right) ^{\widetilde{\varepsilon }%
_{0}}\left( 1-e^{-\alpha r}\right) ^{\nu },\text{ }\widetilde{\varepsilon }%
_{0}>0,\text{ }\nu \geq 1,
\end{equation}%
where 
\begin{equation}
\widetilde{\varepsilon }_{n=0}=\sqrt{\frac{2\mu }{\hbar ^{2}\alpha ^{2}}%
\left( bc_{0}-E_{0}\right) }=\frac{\mu Ze^{2}}{\hbar ^{2}}\frac{1}{\nu
\alpha }-\frac{\nu }{2},
\end{equation}%
with $\nu $ is defined in Eq. (20) and $N_{0}$ is the normalization constant.

Let us find the eigenfunction for any quantum number $n.$ At first,
considering the boundary conditions%
\begin{equation}
y=\left\{ 
\begin{array}{ccc}
0 & \text{when} & r\rightarrow \infty , \\ 
1 & \text{when} & r\rightarrow 0,%
\end{array}%
\right.
\end{equation}%
with $R(y)\rightarrow 0,$ based on Eq. (31), we may define a more general
radial eigenfunctions, valid for any quantum number $n,$ of the form:%
\begin{equation}
R(y)=y^{\widetilde{\varepsilon }_{n,l}}\left( 1-y\right) ^{\nu }F(y),\text{ }%
y=e^{-\alpha r},\text{ }\widetilde{\varepsilon }_{n,l}>0,\text{ }\nu \geq 1,
\end{equation}%
satisfying the boundary conditions in Eq. (33), where%
\begin{equation}
\widetilde{\varepsilon }_{n,l}=\frac{\mu Ze^{2}}{\hbar ^{2}}\frac{1}{\left(
n+\nu \right) \alpha }-\frac{n+\nu }{2}>0.
\end{equation}%
Substituting Eq. (34) into Eq. (8) leads to the following hypergeometric
equation%
\begin{equation}
y\left( 1-y\right) F^{\prime \prime }(y)+\left[ 1+2\widetilde{\varepsilon }%
_{n,l}-\left( 1+2\widetilde{\varepsilon }_{n,l}+2\nu \right) y\right]
F{}^{\prime }(y)-\left[ \nu \left( \nu +2\widetilde{\varepsilon }%
_{n,l}\right) -\frac{2\mu Ze^{2}}{\hbar ^{2}\alpha }\right] F(y)=0,
\end{equation}%
whose solutions are the hypergeometric functions%
\begin{equation}
F(y)=%
\begin{array}{c}
_{2}F_{1}%
\end{array}%
\left( A,B;C;y\right) =\frac{\Gamma (C)}{\Gamma (A)\Gamma (B)}%
\dsum\limits_{k=0}^{\infty }\frac{\Gamma (A+k)\Gamma (B+k)}{\Gamma (C+k)}%
\frac{y^{k}}{k!},
\end{equation}%
where%
\begin{equation*}
A=\widetilde{\varepsilon }_{n,l}+\nu -\sqrt{\widetilde{\varepsilon }%
_{n,l}^{2}+\frac{2\mu Ze^{2}}{\hbar ^{2}\alpha }}=-n,
\end{equation*}%
\begin{equation}
B=\widetilde{\varepsilon }_{n,l}+\nu +\sqrt{\widetilde{\varepsilon }%
_{n,l}^{2}+\frac{2\mu Ze^{2}}{\hbar ^{2}\alpha }},
\end{equation}%
\begin{equation*}
C=1+2\widetilde{\varepsilon }_{n,l}.
\end{equation*}%
By considering the finiteness of the solutions, the quantum condition is
given by%
\begin{equation}
\widetilde{\varepsilon }_{n,l}+\nu -\sqrt{\widetilde{\varepsilon }_{n,l}^{2}+%
\frac{2\mu Ze^{2}}{\hbar ^{2}\alpha }}=-n,\text{ }n=0,1,2,\cdots ,
\end{equation}%
from which we obtain Eq. (25). Now, we may write down the radial wave
functions (34) as%
\begin{equation}
R(r)=\mathcal{N}_{nl}\left( e^{-\alpha r}\right) ^{\widetilde{\varepsilon }%
_{n,l}}\left( 1-e^{-\alpha r}\right) ^{\nu }%
\begin{array}{c}
_{2}F_{1}%
\end{array}%
\left( -n,n+2\left( \widetilde{\varepsilon }_{n,l}+\nu \right) ;1+2%
\widetilde{\varepsilon }_{n,l};e^{-\alpha r}\right) .
\end{equation}%
If we set $n=0$ in Eq. (40)$,$ then we can easily obtain Eq. (31). Finally,
the unnormalized total wave functions are obtained as%
\begin{equation}
\psi _{n,l,m}(r,\Omega _{D})=\mathcal{N}_{nl}r^{-(D-1)/2}\left( e^{-\alpha
r}\right) ^{\widetilde{\varepsilon }_{n,l}}\left( 1-e^{-\alpha r}\right)
^{\nu }%
\begin{array}{c}
_{2}F_{1}%
\end{array}%
(-n,n+2\left( \widetilde{\varepsilon }_{n,l}+\nu \right) ;1+2\widetilde{%
\varepsilon }_{n,l};e^{-\alpha r})Y_{l}^{m}(\Omega _{D}).
\end{equation}%
which is identical to Eq. (42) of Ref. [4] when $D=3.$ Thus, the Jacobi
polynomials can be expressed in terms of the hypergeometric functions [69] 
\begin{equation}
P_{n}^{\left( A,B\right) }(1-2x)=\frac{\Gamma \left( n+1+A\right) }{n!\Gamma
\left( 1+A\right) }%
\begin{array}{c}
_{2}F_{1}%
\end{array}%
(-n,n+A+B+1;A+1;x).
\end{equation}%
The hypergeometric function $_{2}F_{1}(A,B;C;x)$ is a special case of the
generalized hypergeometric function [69,70]%
\begin{equation}
_{p}F_{q}(\alpha _{1},\alpha _{2},\cdots ,\alpha _{p};\beta _{1},\beta
_{1},\cdots ,\beta _{q};x)=\dsum\limits_{k=0}^{\infty }\frac{\left( \alpha
_{1}\right) _{k}\left( \alpha _{2}\right) _{k}\cdots \left( \alpha
_{p}\right) }{\left( \beta _{1}\right) _{k}\left( \beta _{2}\right)
_{k}\cdots \left( \beta _{q}\right) }\frac{x^{k}}{k!},
\end{equation}%
where the Pochhammer symbol is defined by $\left( y\right) _{k}=\Gamma
(y+k)/\Gamma (y).$

Let us find the normalization constant. Introducing the change of parameters 
$y(r)=e^{-\alpha r}$ and making use of Eq. (41), with the help of Eq. (42),
we are able to express the normalization condition $\int_{0}^{\infty
}R(r)^{2}dr=1$ as%
\begin{equation}
\frac{\mathcal{N}_{nl}^{2}}{\alpha }\int_{0}^{1}y^{2\widetilde{\varepsilon }%
_{n,l}-1}(1-y)^{2l+D-1}\left[ P_{n}^{(2\varepsilon _{nl},2l+D-2)}(1-2y)%
\right] ^{2}dy=1.
\end{equation}%
Unfortunately, there is no formula available to calculate this key
integration. Neveretheless, we can find the explicit normalization constant $%
N_{nl}.$ For this purpose, it is not difficult to obtain the results of the
above integral by using the following formulas [69-71],%
\begin{equation}
P_{n}^{(\alpha ,\beta )}(x)=\left( n+\alpha \right) !\left( n+\beta \right)
!\sum\limits_{p=0}^{n}\frac{1}{p!(n+\alpha -p)!\left( \beta +p\right)
!\left( n+p\right) !}\left( \frac{x-1}{2}\right) ^{n-p}\left( \frac{x+1}{2}%
\right) ^{p},
\end{equation}%
and%
\begin{equation}
B(x,y)=\int_{0}^{1}t^{x-1}(1-t)^{y-1}dt=\frac{\Gamma (x)\Gamma (y)}{\Gamma
(x+y)},\text{ }\func{Re}(x),\func{Re}(y)>0.
\end{equation}%
Thus, the normalization constant $\mathcal{N}_{nl}$ is now obtained as%
\begin{equation}
\mathcal{N}_{nl}=\frac{1}{(n+2l+D-2)!\Gamma (2\widetilde{\varepsilon }%
_{n,l}+n+1)}\sqrt{\frac{\alpha \Gamma (2\widetilde{\varepsilon }%
_{n,l}+2n+2l+D+1)}{\Gamma (2\widetilde{\varepsilon }_{n,l}+2n+1)\sum%
\limits_{p,q=0}^{n}\left( f_{p}f_{q}f_{p,q}\right) ^{-1}}},
\end{equation}%
where 
\begin{subequations}
\begin{equation}
f_{p}=(-1)^{p}p!\Gamma (2\widetilde{\varepsilon }_{n,l}+n-p+1)(2l+p+D-2)!%
\left( n+p\right) !,
\end{equation}%
\begin{equation}
f_{q}=(-1)^{q}q!\Gamma (2\varepsilon _{nl}+n-q+1)\left( 2l+D+q-2\right)
!\left( n+q\right) !,
\end{equation}%
\begin{equation}
f_{p,q}=\left( 2l+p+q+D-1\right) !.
\end{equation}%
It is worth noting that one of the disadvantages of the EQR (IQR) approach
is that it cannot get the eigenfunctions of studied potential models. The
estimation shown in this section is only the reverse of logarithmic
derivative to the original Schr\"{o}dinger equation, i.e., $\phi _{0}(r)=%
\frac{d}{dr}\ln (R_{0}(r))$ [see Eq. (30)] [72]. The traditional method is
thus used again.

\section{Conclusions}

In this work, we have applied an alternative method to obtain approximate
energy eigenvalues and eigenfunctions of the $D$-dimensional Schr\"{o}dinger
equation for the Hulth\'{e}n potential with $l\neq 0$ within the improved
approximation scheme for the centrifugal term. The advantage of this method
is that it gives the eigenvalues through the calculation of two integrations
(21) and (22) and solving the resulting algebraic equation. First, we can
easily obtain the quantum correction by only considering the solution of the
ground state of quantum system since it is independent of the number of
nodes of wave function for exactly solvable quantum system. Second, the wave
functions have also been obtained by solving the Riccati equation. The
general expressions obtained for the energy eigenvalues and wave functions
can be easily reduced to the three-dimensional space ($D=3$), $s$-wave ($l=0$%
), the $c_{0}=0$ (usual approximation) cases. The method presented here is a
systematic one, simple, practical.and powerful than the other known methods.
It is worth to extend this method to the solutions of other nonrelativistic
[23,24,34,36-39] and relativistic [35] wave equations with different
potential fields. Finally, it can be also used to deal with many exactly
solvable quantum systems with wide range of potentials as stated by many
authors [23,24,34-39].\ 

\acknowledgments The authors wish to thank the kind referee for the positive
enlightening comments and suggestions which have greatly helped us in making
improvements to the paper. The partial support provided by the Scientific
and Technological Research Council of Turkey is highly appreciated.

\appendix

\section{Integral Formulas}

The following integral formulas are useful during the calculation of the
momentum integral and the quantum correction terms [34,36,68]:\ 

\end{subequations}
\begin{equation}
\int\limits_{r_{A}}^{r_{B}}\frac{r}{\sqrt{(r-r_{A})(r_{B}-r)}}dr=\frac{\pi }{%
2}(r_{A}+r_{B}),
\end{equation}

\begin{equation}
\int\limits_{r_{A}}^{r_{B}}\frac{1}{r\sqrt{(r-r_{A})(r_{B}-r)}}dr=\frac{\pi 
}{\sqrt{r_{A}r_{B}}},
\end{equation}%
\begin{equation}
\int\limits_{r_{A}}^{r_{B}}\frac{1}{\sqrt{(r-r_{A})(r_{B}-r)}}dr=\pi ,
\end{equation}%
\begin{equation}
\int\limits_{r_{A}}^{r_{B}}\frac{1}{r}\sqrt{(r-r_{A})(r_{B}-r)}dr=\pi \left[ 
\frac{1}{2}(r_{A}+r_{B})-\sqrt{r_{A}r_{B}}\right] ,.
\end{equation}%
\begin{equation}
\int\limits_{r_{A}}^{r_{B}}\frac{1}{(a+br)\sqrt{(r-r_{A})(r_{B}-r)}}dr=\frac{%
\pi }{\sqrt{(a+br_{A})(a+br_{B})}},\text{ }r_{B}>r_{A}>0.
\end{equation}

\newpage

{\normalsize 
}


\begin{thebibliography}{99}
\bibitem{1} L. Hulth\'{e}n, Ark. Mat. Astron. Fys. A 28, 5 (1942).

\bibitem{2} S. Fl\"{u}gge, Practical Quantum Mechanics (Springer, Berlin,
1974).

\bibitem{3} S. M. Ikhdair and R. Sever, Phys. Scr. 79, 035002 (2009).

\bibitem{4} S. M. Ikhdair, Eur. Phys. J. A 39, 307 (2009).

\bibitem{5} S. M. Ikhdair and R. Sever, Appl. Math. Comp. 216, 911 (2010).

\bibitem{6} W. C. Qiang, R. S. Zhou and Y. Gao, Phys. Lett. A 371, 201
(2007).

\bibitem{7} R. L. Hall, J. Phys. A: Math. Gen. 25, 1373 (1992).

\bibitem{8} B. Roy and R. Roychoudhury, J. Phys. A 20, 3051 (1987).

\bibitem{9} A. Bechler and W. B\"{u}hring, J. Phys. B 21, 817 (1988).

\bibitem{10} Y. P. Varshni, Phys. Rev. A 41, 4682 (1990).

\bibitem{11} Shishan Dong, J. Garcia-Ravelo and Shi-Hai Dong, Phys. Scr. 76,
393 (2007).

\bibitem{12} S.-H. Dong, Int. J. Quan. Chem. 109 (4), 701 (2009).

\bibitem{13} S.-H. Dong, W.-C. Qiang, G.-H. Sun and V.B. Bezerra, J. Phys. A
40, 10535 (2007); G.-F. Wei, S.-H. Dong and V.B. Bezerra, Int. J. Mod. Phys.
24, 161 (2009); W.-C. Qiang, J.-Y. Wu and S.-H. Dong, Phys. Scr. 79, 065011
(2009).

\bibitem{14} S. M. Ikhdair, J. Math. Phys. 51 (2), 023525 (2010); X.-Y. Gu,
S.-H. Dong and Z.-Q. Ma, J. Phys. A 42, 035303 (2009).

\bibitem{15} W.-C. Qiang and S.-H. Dong, Phys. Scr. 79, 045004 (2009); W.-C.
Qiang and S.-H. Dong, Phys. Lett. A 368, 13 (2007); S.M. Ikhdair, On the
bound-state solutions of the Manning-Rosen potential including improved
approximation to the orbital centrifugal term, to be published in Phys. Scr.

\bibitem{16} F. Cooper, A. Khare and U. Sukhatme, Phys. Rep. 251, 267 (1995).

\bibitem{17} D. A. Morales, Chem. Phys. Lett. 394, 68 (2004).

\bibitem{18} A. F. Nikiforov and V. B. Uvarov, Special Functions of
Mathematical Physics; Birkhaauser: Basel, 1988.

\bibitem{19} S. M. Ikhdair and R. Sever, Appl. Math. Comp. 216, 545 (2010).

\bibitem{20} O. Bayrak, G. Kocak and I. Boztosun, J. Phys. A 39, 11521
(2006).

\bibitem{21} J. B. Killingbeck, A. Grosjean and G. Jolicard, J. Chem. Phys.
116, 447 (2002).

\bibitem{22} M. Bag, M. M. Panja, and R. Dutt, Phys. Rev. A 46, 6059 (1992).

\bibitem{23} Z. Q. Ma and B. W. Xu, Europhys Lett. 69, 685 (2005).

\bibitem{24} Z. Q. Ma and B. W. Xu, Int. J. Mod. Phys. E 14, 599 (2005).

\bibitem{25} B. G\"{o}n\"{u}l, K. Koksal and E. Bak\i r, Phys. Scr. 73, 279
(2006).

\bibitem{26} S. M. Ikhdair and R. Sever, Int. J. Mod. Phys. A 21, 6465
(2006).

\bibitem{27} S. M. Ikhdair and R. Sever, J. Mol. Struc.-Theochem 809, 103
(2007).

\bibitem{28} S. M. Ikhdair and R. Sever, J. Math. Chem. 41, 329 (2007).

\bibitem{29} S. M. Ikhdair and R. Sever, J. Math. Chem. 41, 343 (2007).

\bibitem{30} S. M. Ikhdair and R. Sever, J. Mol. Struc.-Theochem 806, 155
(2007).

\bibitem{31} S. H. Dong, Phys. Scr. 65, 289 (2002).

\bibitem{32} S. H. Dong, Factorization Method in Quantum Mechanics
(Springer, Netherlands, 2007).

\bibitem{33} I. Nasser \textit{et al}, J. Phys. B: At. Mol. Opt. Phys. 40,
4245 (2007).

\bibitem{34} W. -C. Qiang and S. -H. Dong, Phys. Lett. A 363, 169 (2007).

\bibitem{35} W. -C. Qiang, R. -S. Zhou and Y. Gao, J. Phys. A: Math. Theor.
40, 1677 (2007).

\bibitem{36} S. M. Ikhdair and R. Sever, J. Math. Chem. 45 (4), 1137 (2009).

\bibitem{37} Z. -Q. Ma, A. Gonzalez-Cisneros, B. -W. Xu and S. -H. Dong,
Phys. Lett. A 371, 180 (2007).

\bibitem{38} S. -H. Dong and A. Gonzalez-Cisneros, Ann. Phys. 323, 1136
(2008).

\bibitem{39} X. -Y. Gu and J. -Q. Sun, J. Math. Phys. 51 (2), 022106 (2010).

\bibitem{40} S. M. Ikhdair and R. Sever, Int. J. Mod. Phys. A 25 (20), 3941
(2010).

\bibitem{41} R. L. Greene and C. Aldrich, Phys. Rev. A 14, 2363 (1976).

\bibitem{42} S. M. Ikhdair and R. Sever, J. Math. Chem. 42, 461 (2007).

\bibitem{43} N. Saad, Phys. Scr. 76, 623 (2007).

\bibitem{44} W.-C. Qiang and S.-H. Dong, EPL 89, 10003 (2010).

\bibitem{45} F.A. Serrano, X.-Y. Gu and S.-H. Dong, J. Math. Phys. 51,
082103 (2010).

\bibitem{46} C. Yin, Z. Cao and Q. Shen, Ann. Phys. 325, 528 (2010).

\bibitem{47} G.-F. Wei and S.-H. Dong, Phys. Lett. A 373, 2428 (2009).

\bibitem{48} G.-F. Wei and S.-H. Dong, EPL 87, 40004 (2009).

\bibitem{49} G.-F. Wei and S.-H. Dong, Phys. Lett. A 373, 49 (2008).

\bibitem{50} G.-F. Wei and S.-H. Dong, Phys. Lett. B 686, 288 (2010).

\bibitem{51} F. Dominguez-Adame, Phys. Lett. A 136, 175 (1989).

\bibitem{52} L. Chetouani, L. Guechi, A. Lecheheb, T. F. Hammann and A.
Messouber, Physics A 234, 529 (1996).

\bibitem{53} B. Talukdar, A. Yunus and M. R. Amin, Phys. Lett. A 141, 326
(1989).

\bibitem{54} Z.-Y. Chen, M. Li , C.-S. Jia, Mod. Phys. Lett. A 24 (23), 1863
(2009).

\bibitem{55} Y.-F. Diao, L.-Z. Yi, T. Chen, C.-S. Jia, Mod. Phys. Lett. B
23, 2269 (2009).

\bibitem{56} T. Chen, J.-Y. Liu, C.-S. Jia, Phys. Scr. 79, 055002 (2009).

\bibitem{57} T. Chen, Y.-F. Diao, C.-S. Jia, Phys. Scr. 79, 065014 (2009).

\bibitem{58} C.-S. Jia, T. Chen, L.-G. Cui, Phys. Lett. A 373 (18-19), 1621
(2009).

\bibitem{59} C.-S. Jia, J.-Y. Liu and P.-Q. Wang, Phys. Lett. A 372, 4779
(2008).\ 

\bibitem{60} B. G\"{o}n\"{u}l, O. \"{O}zer, Y. Can\c{c}elik and M. Kocak,
Phys. Lett. A 275, 238 (2000).

\bibitem{61} B. G\"{o}n\"{u}l, Chin. Phys. Lett. 21, 1685 (2004).

\bibitem{62} R. Dutt, K. Choudhury and Y. P. Varshni, J. Phys. A: Math. Gen.
18, 1379 (1985).

\bibitem{63} T. Xu, Z. -Q. Cao, Y. -C. Ou, Q. -S. Shen and G. -L. Zhu, Chin.
Phys. 15, 1172 (2006).

\bibitem{64} E. D Filho and R. M. Ricotta, Mod. Phys. Lett. A 10, 1613
(1995).

\bibitem{65} V. B. Mandelzweig, Ann. Phys. 321, 2810 (2006).

\bibitem{66} S. M. Ikhdair, Int. J. Mod. Phys. C 20 (1), 25 (2009).

\bibitem{67} C.-S. Jia, J.-Y. Liu and P.-Q. Wang, Phys. Lett. 372, 4779
(2008).

\bibitem{68} S.-H. Dong, D. Morales and J. Garcia-Ravelo, Int. J. Mod. Phys.
E 16, 189 (2007).

\bibitem{69} I.S. Gradshteyn and I.M Ryzhik, Tables of Integrals, Series,
and Products, 5th edn (New York, Academic, 1994).

\bibitem{70} G. Sezgo, Orthogonal Polynomials, (American Mathematical
Society, New York, 1939).

\bibitem{71} W. Magnus, F. Oberhettinger and R.P. Soni, Formulas and
Theorems for the Special Function of Mathematical Physics, 3rd Ed., (Berlin,
\ Springer, 1966).

\bibitem{72} X.-Y. Gu, M. Zhang and J.-Q. Sun, Chin. J. Phys. 48, 222 (2010).
\end{thebibliography}
\end{document}